\documentclass[aps,prl,twocolumn,showpacs,showkeys,superscriptaddress]{revtex4-2}

\usepackage{graphicx}
\usepackage{rotating}
\usepackage{amssymb}
\usepackage{amsmath}
\usepackage{epsfig}
\usepackage[normalem]{ulem}
\usepackage{bm}
\usepackage{color}
\usepackage{physics} 
\usepackage{mathtools}
\usepackage{enumerate} 
\usepackage{enumitem}
\usepackage{phfcc}

\phfMakeCommentingCommand[initials={M.R.}]{MR}

\date{\today}

\begin{document}

\title{Dynamic link switching induces stable synchronized states in sparse networks}

\author{Muhittin Cenk Eser}
\thanks{Author to whom correspondence should be addressed: muhittin.eser@fu-berlin.de}
\affiliation{Department of Mathematics and Computer Science, Freie Universit\"at Berlin, Arnimallee 6, 14195 Berlin, Germany}

\author{Everton S. Medeiros}
\affiliation{S\~ao Paulo State University (UNESP), Institute of Geosciences and Exact Sciences, Avenida 24A 1515, 13506-900 Rio Claro, S\~ao Paulo, Brazil}

\author{Mustafa Riza}
\affiliation{Department of Physics, Eastern Mediterranean University, 99628 Famagusta, North Cyprus, via Mersin 10, Turkey}

\author{Maximilian Engel}
\affiliation{Department of Mathematics and Computer Science, Freie Universit\"at Berlin, Arnimallee 6, 14195 Berlin, Germany}
\affiliation{KdV Institute, University of Amsterdam, Science Park 105-107, 1098 XG Amsterdam, The Netherlands}

\begin{abstract}
The flow of information in networked systems composed of multiple interacting elements strongly depends on the level of connectivity among these elements. Sparse connectivity often hinders the emergence of states in which information is globally shared, such as fully synchronized states. In this context, dynamically switching existing network links among system elements can facilitate the onset of synchronization. Here, we address this problem in a double-layer network of FitzHugh–Nagumo oscillators with sparse inter-layer connectivity at fixed density. We show that dynamically switching the existing cross-layer links induces inter-layer synchronization, with a clear dependence on the switching time. In agreement with intuition, shorter switching times suppress large deviations between temporally connected oscillators and more effectively promote synchronization; crucially, this effect persists even when each isolated layer is chaotic. Chaos at the layer level is verified by a strictly positive largest Lyapunov exponent, confirming that synchrony is induced by switching rather than by periodic dynamics. For a minimal double-layer system, we emulate switching using smooth square waves and compute the master stability function (MSF), which is in agreement with direct numerical simulations and delineates the stability regions in parameter space.
\end{abstract}
\maketitle

\section{Introduction}

Many technological systems depend on efficient methods of information exchange between their various components in order to function properly. Examples include wireless sensor networks, where spatially distributed sensors communicate with central control systems \cite{Raghavendra2006}; smart grid systems, where data from distributed energy resources, substations, and smart meters converge at control centers \cite{Kabalci2019}; and autonomous vehicle networks, where vehicles and infrastructure continuously exchange data to manage traffic, prevent collisions, and coordinate movement \cite{Wang2019}. Interestingly, to minimize energy consumption at the component level and optimize bandwidth usage with the central control system, communication between components and the control center occurs intermittently, activated only when necessary to update the system. This intermittent exchange of information raises questions about the thresholds for the optimal duration that a connection must remain active to ensure consensus of information across the entire system.

In the theoretical study of dynamical systems, the effects of intermittent information exchange have been extensively explored in the context of synchronization in networks of coupled nonlinear systems. For instance, in a network of chaotic oscillators, it has been shown that coupling the oscillators for only a fraction of their state space can lead to synchronization, whereas permanent coupling within the full state space does not \cite{Timme2015,Timme2016}.
Ref.~\cite{Chen2009} investigates the stability of synchronized states in networks where all links are switched on and off with specified time scales. Additionally, Ref.~\cite{Sansan2018} demonstrates that chaotic oscillators periodically coupled using a sinusoidal function can achieve enhanced synchronization by appropriately adjusting the coupling frequency. Interestingly, periodic coupling may also be detrimental to synchronization \cite{Li2018,Baumann2020}. In Ref.~\cite{Njougouo2020}, two types of oscillators are intermittently coupled when the distances between specific state variables of certain oscillators fall below a predefined threshold. This approach revealed convergence to synchronization that depends on the chosen threshold. Furthermore, in a double-layer network of FitzHugh-Nagumo (FHN) oscillators, randomly switching the connections between oscillators in different layers induces an abrupt transition to inter-layer synchronization as the number of connections increases \cite{Cenk2021}. Finally, beyond the aforementioned studies, a wide range of other topological time dependencies have been investigated to enhance synchronization in both single-layer networks \cite{Boccaletti2006,Zhou2006,Kohar2014,Zhou2019,Ghosh2022,Kojakhmetov2024} and multilayer networks \cite{Rakshit2018,Rakshit2019,Berner2020,Vadivasova2020}.

Despite advancements in the literature, little is known about the optimal time interval during which connections between certain elements should remain active before switching to different elements, ensuring that the entire system remains synchronized. Specifically, the following questions arise: given a limited number of connections, is it more advantageous for synchronization to maintain the connection between certain elements for a longer duration, or is it beneficial to shorten the connection duration and switch to different elements? Furthermore, are the synchronized states resulting from such switching asymptotically stable?

In this work, we investigate these questions within the framework of synchronization in a double-layer network of FitzHugh-Nagumo oscillators. Each isolated layer consists of two distinct populations of FHN oscillators diffusively coupled to their nearest neighbors via intra-layer connections. The individual dynamics of each layer is chaotic, with the oscillators within each layer remaining desynchronized. Additionally, the two identical layers are sparsely interconnected through inter-layer links, which diffusively couple pairs of oscillators from each layer (mirrored oscillators). The layers are initially desynchronized with respect to one another.

In this context, we periodically switch the existing inter-layer links at a fixed time interval, referred to as the switching time. At each switching event, the existing inter-layer links are randomly reassigned between corresponding oscillators in the two layers. We observe that shortening the switching time, thereby allowing inter-layer links to sweep through a larger number of oscillators across the layers, is more effective for achieving synchronization than maintaining the same pair of mirror oscillators connected for longer time intervals. This is demonstrated by analyzing the system's time evolution under different switching times. Furthermore, the effectiveness of the switching procedure is evaluated for increasing layer sizes, highlighting its performance to establish synchronization across different switching times. Additionally, we investigate the time to synchrony for different switching times and population sizes, while keeping the ratio between the interlayer links and the population size constant, complying with the sparse network requirement. Also, the minimum switching time as a function of the population while keeping the ratio of the number of interlayer links to the population size constant for desynchrony is another approach to show, quantitatively, the synchronization dependency on the switching time and the population size.  Finally, to examine the asymptotic stability of the synchronized states induced by the switching procedure, we analyze a minimal double-layer system consisting of two FHN oscillators per layer, interconnected by a single inter-layer link. To address the eventual discontinuities implied in the switching procedure, we model the switching of the inter-layer link using a smooth square-wave function. Using this approach, we apply the master stability function (MSF) formalism to determine the interval of switching times and coupling intensities for which the synchronized states are stable.

\section{Model: A double-layer network of FitzHugh-Nagumo oscillators}

\begin{figure}[!htp]
    \centering
    \includegraphics[width=\columnwidth]
    {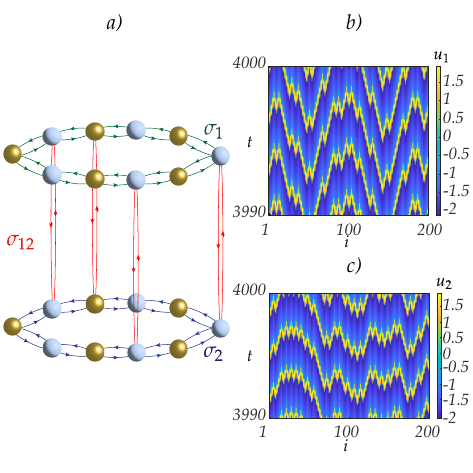}
    \caption{(a) Schematic of a double-layer network illustrating the two distinct populations of FHN oscillators in its nodes, represented in blue and gold. The constants $\sigma_1$ and $\sigma_2$ denote the intra-layer coupling intensities, while $\sigma_{12}$ represents the inter-layer coupling intensity. (b) Spatiotemporal dynamics of the activator variable $u_{1j}$ for the FHN oscillators in layer $1$. (c) Spatiotemporal dynamics of the activator variable $u_{2j}$ for the FHN oscillators in layer $2$. Parameters are $\varepsilon = 0.05$, $a_{ij}=0.87$ (even $j$), $a_{ij}=0.97$ (odd $j$), $\phi = \frac{\pi}{2} - 0.1$, $\sigma_{12} = 0$, $\sigma_1 = \sigma_2 = 0.1$, and $N=200$. }
    \label{fig:schematic_time_evolution}
\end{figure}

To investigate the appropriate time intervals during which a network connection must remain active to ensure complete network synchronization, we consider a double-layer network with nodes composed of FitzHugh-Nagumo (FHN) oscillators. Each layer comprises $N$ FHN oscillators diffusively coupled to their nearest neighbors under periodic boundary conditions, forming ring topologies. The two layers are interconnected through a specified number of inter-layer links $N_{IL}$, which connect nodes positioned identically in each layer, referred to as mirror nodes (see the schematic picture in Fig.~\ref{fig:schematic_time_evolution}(a)). Similar double- and multilayer networks have recently been employed to explore a variety of dynamical phenomena (e.g., \cite{Medeiros2021,Semenov2022,Semenov2023,Zakharova2025}); see also Ref. \cite{Wu2024} for a comprehensive review. Similar multilayer arrangements have been extensively studied in both deterministic \cite{Omelchenko2013, Rybalova2019,Mikhaylenko2019,Rybalova2021,Schulen2021} and stochastic systems \cite{Zakharova2017,Masoliver2017,Semenova2018,Masoliver2021}. Here, we consider a fixed partial inter-layer link density $N_{IL}/N=0.25$. We emphasize that the total number of possible inter-layer links is $N(N-1)/2$. Since $N_{IL} \ll N(N-1)/2$, the inter-layer connectivity considered here is indeed very sparse. Additionally, in our double-layer network, the positions of the inter-layer links are time-dependent. These links remain active between a set of mirror nodes for a finite time interval before randomly switching to a different set. This interval is defined as the switching time, $T_{\rm swt}$. To provide a more precise description of this system, we define its spatiotemporal dynamics using the following equations:
\begin{multline}
\dv{}{t}\begin{pmatrix} \mathbf{L}_1 \\ \mathbf{L}_2 \end{pmatrix} = \begin{pmatrix} \vb{F}(\mathbf{L}_1) \\ \vb{F}(\mathbf{L}_2) \end{pmatrix} + \begin{pmatrix} \sigma_1 \left(\mathcal{L} \otimes \mathcal{H}\right) \mathbf{L}_1 \\ \sigma_2 \left(\mathcal{L} \otimes \mathcal{H}\right) \mathbf{L}_2 \end{pmatrix} + \\ + \sigma_{12} \left(\mathcal{L}^I \otimes \Gamma(t)\right) \begin{pmatrix} \mathbf{L}_1 \\ \mathbf{L}_2 \end{pmatrix},
\label{eqn:model}
\end{multline}
where $\mathbf{L}_i = \left[u_{i1}, v_{i1}, \dots, u_{iN}, v_{iN} \right]^\top$ represents the state vector of the $i$-th layer ($i = 1, 2$), which includes the activator $u_{ij}$ and inhibitor $v_{ij}$ of the $j$-th ($j=1,2,\dots,N$) node in the $i$-th layer. The dynamics of the FHN oscillators \cite{FitzHugh1961,Nagumo1962} at the nodes of each layer are governed by:
\begin{equation}
\vb{F}(\mathbf{L}_i) = \begin{pmatrix}
\frac{1}{\varepsilon} \left(u_{i1} - \frac{u_{i1}^3}{3} - v_{i1}\right) \\
u_{i1} + a_{i1} \\
\vdots \\
\frac{1}{\varepsilon} \left(u_{iN} - \frac{u_{iN}^3}{3} - v_{iN}\right) \\
u_{iN} + a_{iN}
\end{pmatrix},
\label{eqn:FHN}
\end{equation}
where $\varepsilon = 0.05$ is a control parameter that specifies the time scale separation between the fast activator $u_{ij}$ and the slow inhibitor $v_{ij}$. The excitability threshold for the parameter $a_{ij}$ determines whether the FHN oscillator operates in its excitable ($|a_{ij}| > 1$) or oscillatory ($|a_{ij}| < 1$) regime. All oscillators in this work operate in the oscillatory regime; however, each layer consists of two distinct populations of FHN oscillators, differentiated by their respective values of $a_{ij}$.  
These oscillators are alternately distributed throughout the ring topology: for odd-indexed oscillators ($j=1, 3, 5, \dots$), the value is $a_{ij}=0.87$, while for even-indexed oscillators ($j=2, 4, 6, \dots$), the value is $a_{ij}=0.97$  (see golden and blue nodes in Fig.~\ref{fig:schematic_time_evolution}(a)). This intra-layer heterogeneity reduces the likelihood of synchronization among the FHN oscillators within each layer. However, the two layers considered are identical, and each pair of mirror nodes hosts FHN oscillators of the same type, as illustrated in Fig.~\ref{fig:schematic_time_evolution}(a).

The parameters $\sigma_1$ and $\sigma_2$ in Eq.~(\ref{eqn:model}) regulate the coupling intensity among the FHN oscillators within layers $1$ and $2$, respectively. Additionally, the intra-layer nearest-neighbor coupling structure is defined by the zero row sum Laplacian matrix $\mathcal{L}$ as given by:
\begin{equation}
\mathcal{L} = \begin{pmatrix}
-2 & 1 & 0 & 0&0&\cdots 0 & 1 \\
1 & -2 & 1 & 0 &0& \cdots& 0 \\
0 & 1 & -2 &1 & 0&  \cdots& 0 \\
0 & 0&1 & -2 &1 & 0\cdots & 0 \\
\vdots &\vdots &\ddots & \ddots & \ddots & \ddots &\vdots\\
0 &\cdots&0&1&-2&1&0\\
0 &\cdots&\cdots&0&1&-2&1\\
1 & 0 & \cdots &\cdots&0& 1 & -2
\end{pmatrix}.
\label{eqn:laplacian_intra}
\end{equation}

The intra-layer coupling matrix $\mathcal{H}$ accounts for the cross-influence between the activator ($u$) and the inhibitor ($v$) \cite{Schulen2022}, an effect governed by a rotational coupling matrix \cite{Omelchenko2013}:
\[
\mathcal{H} = \begin{pmatrix}
\cos{\phi} & \sin{\phi} \\
-\sin{\phi} & \cos{\phi}
\end{pmatrix},
\]
where the phase angle $\phi \in [-\pi, +\pi)$ plays a crucial role in shaping the dynamic behavior within each isolated layer \cite{Rybalova2019,Cenk2021}. Following the cited literature, we set the cross-coupling parameter to $\phi = \frac{\pi}{2} - 0.1$, a value that allows desynchronized behavior within each layer.

The coupling between the layers (inter-layer coupling) is diffusive, with the intensity regulated by $\sigma_{12}$ and the structure defined by the inter-layer Laplacian $\mathcal{L}^{I}$. The inter-layer coupling matrix $\Gamma(t)$ governs the coupling exclusively between the activator variable $u$ of mirror oscillators. These matrices, $\mathcal{L}^{I}$ and $\Gamma(t)$, are defined as follows:
\begin{eqnarray*}
\mathcal{L}^I &=& \begin{pmatrix}
-1 & 1 \\
1 & -1
\end{pmatrix}
\qquad\text{ and }\\
\Gamma(t) &=& 
 \begin{pmatrix}
\gamma_1(t) & 0 & 0 & \cdots & 0 & 0 \\
0 & 0 & 0 & \cdots & 0 & 0 \\
0 & 0 & \gamma_{2}(t) & \cdots & 0 & 0 \\
\vdots & \vdots & \vdots & \ddots & \vdots \\
0 & 0 & 0 & \cdots & \gamma_{2N}(t) & 0\\
0 & 0 & 0 & \cdots & 0 & 0
\end{pmatrix}.
\end{eqnarray*}
The time dependence $\gamma_{j}(t)$ in the components of $\Gamma(t)$ reflects the switching of inter-layer links at each time interval, i.e., the switching time $T_{swt}$. If an inter-layer link exists for oscillator $j$, then $\gamma_{j}(t) = 1$; otherwise, $\gamma_{j}(t) = 0$. In contrast to \cite{Rakshit2019}, where a time-dependent rewiring is introduced implicitly, we apply a real time-dependent wiring, where we can define the rewireing frequency explicitly. Finally, the FHN oscillators in each layer are initialized with initial conditions (ICs) randomly selected from the intervals $u_{ij}(0) \in \left[-2.0, 2.0\right]$ and $v_{ij}(0) \in \left[-2.0, 2.0\right]$. In this way, both the FHN oscillators within the layers and the layers themselves are desynchronized at the initial time.

Following these definitions, we numerically integrate Eq.~(\ref{eqn:model}) to analyze the spatiotemporal dynamics of our double-layer network. As a first step, we consider each layer in isolation by setting the inter-layer coupling intensity to zero, i.e., $\sigma_{12} = 0$. Meanwhile, the intra-layer coupling intensities are fixed at $\sigma_1 = \sigma_2 = 0.1$. Each layer contains $N = 200$ FHN oscillators. Figures~\ref{fig:schematic_time_evolution}(b) and \ref{fig:schematic_time_evolution}(c) illustrate the spatiotemporal dynamics of the activator $u_{ij}$ for each FHN oscillator $j$ ($j = 1, \dots, N$) in layers $i = 1$ and $i = 2$, respectively. These figures highlight the asynchronous dynamics both among the FHN oscillators within each layer and between the layers.

\begin{figure}[!htp]
    \centering
    \includegraphics[width=\columnwidth]{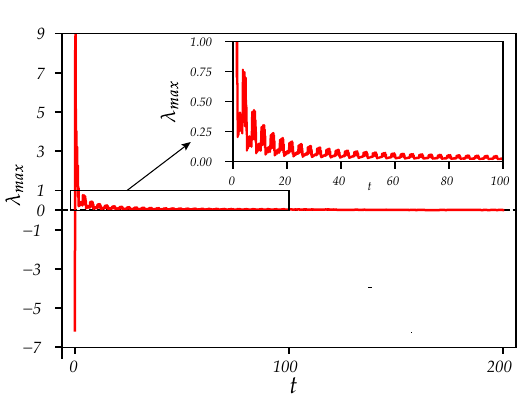}
    \caption{The largest Lyapunov exponent $\hat{\lambda}_{\max}(t)$ as a function of time for an isolated layer ($\sigma_{12}=0$), computed via the Benettin method (with periodic renormalization of tangent vectors), converging to $\lambda_{\max} \approx 0.04$. The simulation parameters are $\varepsilon = 0.05$, $a_{\text{odd}} = 0.87$, $a_{\text{even}} = 0.97$, $\phi = \frac{\pi}{2} - 0.1$, $\sigma_1 = \sigma_2 = 0.1$, and $N = 200$, with local coupling radius $r = 1$. A positive $\lambda_{\max}$ confirms the presence of chaotic dynamics within each layer.}
\label{fig:mle_layer}
\end{figure}

Due to the distinct set of initial conditions (ICs) assigned to each identical layer, their spatiotemporal dynamics, while uncoupled, as shown in Fig.~\ref{fig:schematic_time_evolution}, are desynchronized. When the layers are coupled through the diffusive inter-layer coupling defined in Eq.~(\ref{eqn:model}), synchronization may occur, typically depending on topological properties such as the coupling connectivity and strength or, as we demonstrate here, on the switching time. However, the intrinsic dynamical features of the layers are also decisive for the onset of synchronization. In particular, when the layer dynamics exhibit a limit cycle, synchronization is much easier to achieve because limit cycles possess strong transverse contraction, so that the inter-layer coupling only needs to align the phases. In contrast, when the layer dynamics are chaotic and characterized by positive Lyapunov exponents, synchronization requires the coupling to continuously suppress intrinsic exponential divergence, leading to narrower stability regions in the space of topological parameters (coupling intensity and connectivity). To confirm that, for the parameters considered here, the uncoupled layers are indeed chaotic, thereby posing greater challenges for the switching-time strategy, we compute in Fig.~\ref{fig:mle_layer} the largest Lyapunov exponent of the identical layers using the Benettin method \cite{benettin1980lyapunov}.

\section{Results}

\subsection{Link-switching-induced inter-layer synchronization}

We now demonstrate the role of switching the existing inter-layer links in establishing synchronization between the layers. Naturally, the time interval during which the links remain active before switching, referred to as the switching time $T_{swt}$, plays a central role in this process. To explore this role, we first analyze the time evolution of the double-layer system for different values of $T_{swt}$ in a network with $N = 400$ FHN oscillators per layer and only $N_{IL} = 100$ inter-layer links connecting the mirror nodes between the layers.

The parameters used for the figures are all common and as following: $\varepsilon = 0.05$, $a_{ij}=0.87$ (even $j$), $a_{ij}=0.97$ (odd $j$), $\phi = \frac{\pi}{2} - 0.1$, $\sigma_1 = \sigma_2 = 0.1$.

\begin{figure}[!htp]
    \centering
    \includegraphics[width=\columnwidth]{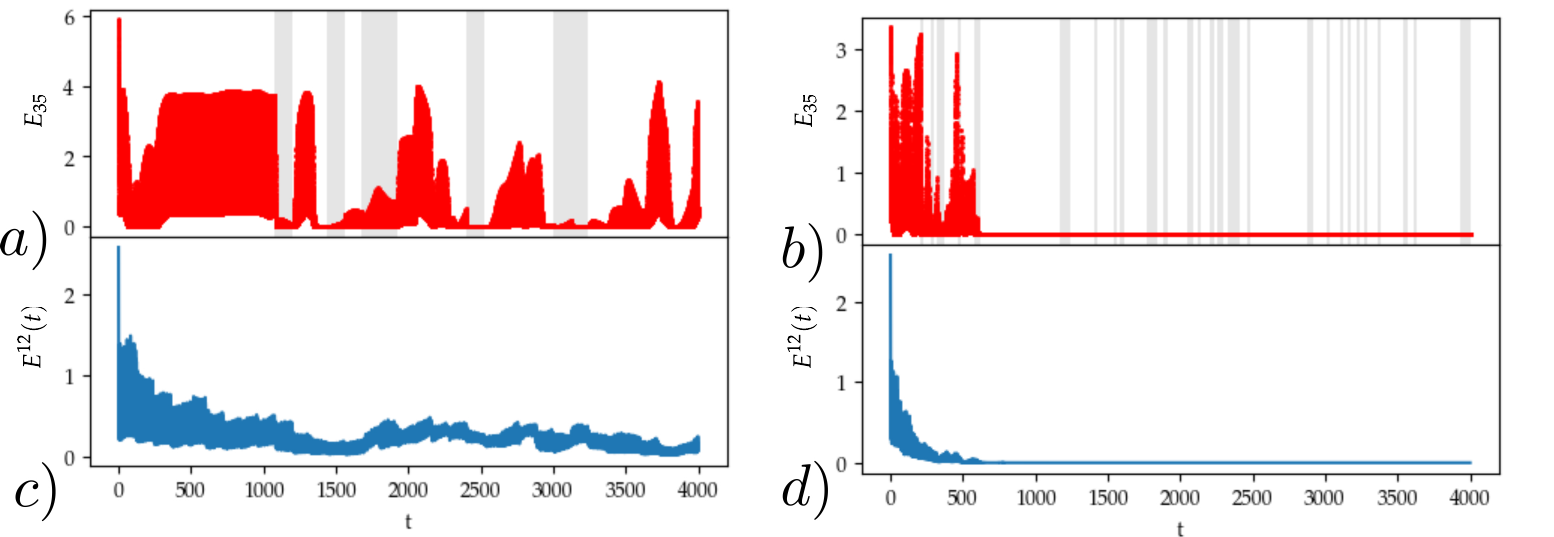}
    \caption{Time evolution of the Euclidean distance $E_{35}$ between mirror nodes number $35$ is marked in red for the switching time a) $T_{swt}=120$ and b) $T_{swt}=23$. The time intervals marked in gray indicate the periods in which mirror nodes 35 are connected, while in white are time intervals in which these nodes are disconnected. In c) and d) blue curve shows the time evolution of the interlayer synchronization error $E^{12}(t)$ for $T_{swt}=120$ and $T_{swt}=23$, respectively.    Additionally to the given parameters, $
   \sigma_{12}=0.1$ and $N=400.$} 
    \label{fig:Euclidean_distances}
\end{figure}

In Fig.~\ref{fig:Euclidean_distances}(a), for $T_{swt} = 120$, the red curve represents the time evolution of the Euclidean distance between mirror nodes numbered $35$, defined as $E_{35} = \sqrt{(u_{i=1j=35} - u_{i=2j=35})^2 + (v_{i=1j=35} - v_{i=2j=35})^2}$. In this figure, the time intervals shown in gray correspond to the presence of an inter-layer link between the mirror nodes $35$, while the time intervals shown in white indicate the absence of such a link. We observe that the distance $E_{35}$ exhibits high-amplitude oscillations during the time intervals marked in white, corresponding to the absence of an inter-layer link. In contrast, during the gray intervals, where an inter-layer link is present, $E_{35}$ oscillates with a significantly lower amplitude. Notice that, for $T_{swt} = 120$, immediately after the gray time intervals, regardless of their duration, the amplitude of $E_{35}$ begins to increase as soon as the inter-layer link between mirror nodes $35$ is switched off. This behavior suggests that switching the inter-layer links at every $T_{swt} = 120$ does not result in inter-layer synchronization. On the other hand, in Fig.~\ref{fig:Euclidean_distances}(b), for $T_{swt} = 23$, where the resulting time intervals marked in gray are significantly shorter, we observe a transient phase in which high and low amplitude oscillations alternate depending on the presence or absence of an inter-layer link. However, in this case, right after one gray time interval, the distance $E_{35}$ drops to zero, marking the end of the transient phase and indicating synchronization between the mirror nodes $35$. Finally, we emphasize that, although all links are switched periodically at every $T_{swt}$, a link connecting a specific pair of mirror nodes may be reassigned to the same pair after switching. Consequently, as observed in Fig.~\ref{fig:Euclidean_distances}(a) and Fig.~\ref{fig:Euclidean_distances}(b), the time intervals marked in gray can vary in size. Moreover, since $N_{IL}=0.25N$, a given pair of mirror nodes may not receive an inter-layer link even after several switches, resulting in the time intervals marked in white also varying in size.

Now, we complement the analysis presented in Figs.~\ref{fig:Euclidean_distances}(a) and \ref{fig:Euclidean_distances}(b) by verifying their corresponding behavior with respect to inter-layer synchronization. To achieve this, we first define a measure to estimate the level of inter-layer synchronization at each time instant as:
\begin{eqnarray}
    E^{12}(t) = \frac{1}{N} \| \mathbf{L}_1(t) - \mathbf{L}_2(t) \|,
    \label{eqn:td_error}
\end{eqnarray}
where $E^{12}(t)$ represents the time-dependent inter-layer synchronization error, defined as the Euclidean distance between the state vectors of layer $1$ and layer $2$, normalized by the layer size $N$. Hence, in Fig.~\ref{fig:Euclidean_distances}(c), we plot $E^{12}(t)$ for the large switching time $T_{swt} = 120$, corresponding to the results shown in Fig.~\ref{fig:Euclidean_distances}(a) for $E_{j=35}$. We observe that $E^{12}(t)$ fluctuates around a constant positive value, remaining significantly bounded away from zero (above $10^{-2}$) for all $t \in [0,4000]$, suggesting that the layers do not converge towards synchronization within the observed time window for this value of $T_{swt}$. In other words, the link-switching strategy fails to establish inter-layer synchronization. Conversely, in Fig.~\ref{fig:Euclidean_distances}(d), for the small value of the switching time $T_{swt} = 23$, we observe that $E^{12}(t) \to 0$ after a transient period, indicating the onset of inter-layer synchronization induced by the switching of inter-layer links.

Figure~\ref{fig:Euclidean_distances} illustrates the dynamics of the double-layer system influenced by the switching of inter-layer links. We observe that switching the existing links at every $T_{swt} = 23$ drives the system to inter-layer synchronization. To verify the robustness of this switching process, we evaluate the duration of the transient time required for the link switching to bring the layers into complete inter-layer synchronization.

\begin{figure}[!htp]
    \centering
    \includegraphics[width=\columnwidth]{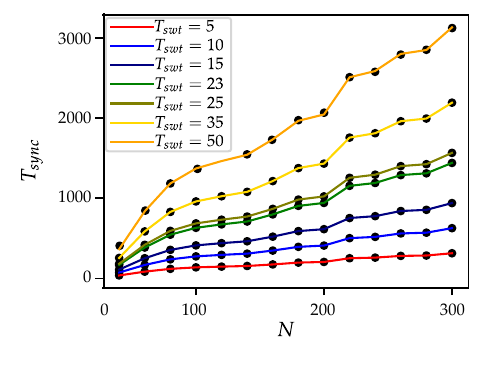}
    \caption{Duration of the transient time, $T_{sync}$, before the onset of complete inter-layer synchronization as a function of the layer size $N$. The ratio $N_{IL}/N=0.25$ is kept constant while changing the population size $N$. The interlayer links are switched randomly. 
    Exemplarily,the transient times to synchronization for $T_{swt}=5,10,15,23,25,35,50$, for the above-mentioned parameters.
   We note the time where the $E^{12}<10^{-4}$.
   }
    \label{fig:transient_system_size}
\end{figure}

In Fig.~\ref{fig:transient_system_size}, we present the duration of the transient time required for the link switching to establish complete inter-layer synchronization, $T_{sync}$, as a function of the layer size $N$ for different switching times $T_{swt} = 5,10,15,23,25,35,50$, while keeping the ratio $N_{IL}/N=0.25$ constant complying with the sparse network requirement. 
The transient time $T_{sync}$ increases monotonically with increasing $N$ for all switching times, demonstrating the effectiveness of the link switching procedure.

\begin{figure}[!htp]
\centering
\includegraphics[width=\columnwidth]{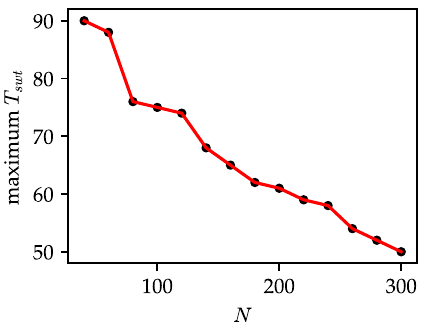}
\caption{This figure illustrates the maximum switching time, where the system still synchronizes under interlayer switching as a function of the layer population, using the abovementioned parameters. The ratio $N_{IL}/N=0.25$ is kept constant while changing the size of the population. The interlayer links are switched randomly.} 
\label{fig:time_desync_size}
\end{figure}

Fig.~\ref{fig:time_desync_size} complements Fig.~\ref{fig:transient_system_size} by illustrating the maximum switching time required for the system to synchronization, based on the number of nodes. For all switching times smaller or equal than the ones illustrated in the figure, the system will converge to the synchronous manifold. 
For switching times greater than those shown in the figure, the system will not synchronize for this population size.

Concerning the dependence of effectiveness of this link-switching strategy on the system parameters, we expect the qualitative effectiveness to persist across different chaotic attractors. Nevertheless, chaotic attractors may differ substantially in the structure of their tangent spaces, including the magnitude and distribution of local contraction and expansion rates. As a result, the quantitative effectiveness of the link-switching strategy may depend on specific features of the underlying chaotic dynamics.

\subsection{Stability analysis of synchronized states}

In the previous section, we observed that with a low density of inter-layer links, the dynamic redistribution of existing links among the FHN oscillators can promote the emergence of synchronous dynamics across the layers. More interestingly, the timing of these redistributions, defined by the switching time $T_{swt}$, plays a crucial role. However, whether the emergent synchronous dynamics is asymptotically stable remains an open question. To address this issue, we perform a transverse stability analysis using the master stability function (MSF) formalism on a simplified system consisting of two FHN oscillators per layer.

To implement the MSF in such a system, we treat each identical layer, comprising two oscillators, as a single higher-dimensional dynamical system, rather than as a network of lower-dimensional systems. Accordingly, we define the function $\mathbf{G}(\mathbf{L}_i) := \mathbf{F}(\mathbf{L}_i) + \sigma_i\,(\mathcal{L}\!\otimes\!\mathcal{H})\,\mathbf{L}_i$, with $i=1,2$ and $\sigma\equiv\sigma_1=\sigma_2$, which encapsulates the local dynamics of the FHN oscillators defined in Eq.~(\ref{eqn:FHN}) and the intra-layer coupling prescribed by the matrices $\mathcal{L}$ and $\mathcal{H}$, together with the parameter $\sigma_i$. Using this approach, Eq.~\eqref{eqn:model} can be rewritten as:
\begin{equation}
\dot{{\bf \;L}}_i = \vb{G}(\mathbf{L}_i)  + \sigma_{12} \left(\mathcal{L}^I \otimes \Gamma(t)\right) \mathbf{L}_i,
\label{eqn:layer}
\end{equation}
where $\mathbf{L}_i$ continue to represent the dynamical state of the entire layer, which now comprises $N=2$ FHN oscillators labeled as $j=1,2$.

The complete synchronized inter-layer state in the double-layer system occurs on a synchronization manifold $\mathbf{S}$ which is defined as $\mathbf{S} = \mathbf{L}_1 = \mathbf{L}_2$. To assess the local stability of $\mathbf{S}$ using the MSF formalism, we introduce infinitesimal perturbations $\delta \mathbf{L}_i$ in all directions of the layer state space. This approach leads to the following variational equation governing the time evolution of $\delta \mathbf{L}_i$:
\begin{equation}
 \dot{\delta{\bf L\;}}_i = \left[ \mathbf{DG}(\mathbf{S}) + \sigma_{12} (\mathcal{L}^I \otimes \mathbf{\Gamma}(t)) \right] \cdot \delta \mathbf{L}_i,
 \label{eqn:variational}
\end{equation}
where $\mathbf{DG}(\mathbf{S})$ represent the Jacobian matrices of the intra-layer dynamics evaluated at the synchronization manifold $\mathbf{S}$. The Laplacian matrix $\mathcal{L}^I$ is diagonalized with real eigenvalues $\lambda_i$ for $i = 1, 2$. Consequently, Eq.~(\ref{eqn:variational}) becomes block diagonalizable, yielding a decoupled variational equation:
\begin{eqnarray}
\delta\dot{\mathbf{\Theta}}_i = \left[ \mathbf{DG}(\mathbf{S}) - \sigma_{12} |\lambda_i| \mathbf{\Gamma}(t)\right] \cdot \delta \mathbf{\Theta}_i,
\label{eqn:variational_decoupled}
\end{eqnarray}
where $\lambda_1 = 0$ and $\lambda_2 = -2$. Since $\lambda_1 = 0$ corresponds to directions tangent to the dynamic flow, it does not contribute to the transverse stability. Therefore, only the eigenvalue $\lambda_2 = -2$ shall be considered in our analysis. Taking this into account in Eq.~(\ref{eqn:variational_decoupled}), we obtain:
\begin{eqnarray}
\delta\dot{\mathbf{\Theta}} = \left[ \mathbf{DG}(\mathbf{S}) - 2\sigma_{12} \mathbf{\Gamma}(t)\right] \cdot \delta \mathbf{\Theta}.
\label{eqn:variational_decoupled_2}
\end{eqnarray}
The largest Lyapunov exponent of the system in Eq.~(\ref{eqn:variational_decoupled_2}) can be represented by the MSF $\Psi$. Since the coupling intensity $\sigma_{12}$ and the switching time $T_{swt}$ are the parameters of interest, we obtain the MSF as function of these parameters $\Psi(\sigma_{12}, T_{swt})$.

\begin{figure}[!htp]
    \centering
    \includegraphics[width=\columnwidth]
    {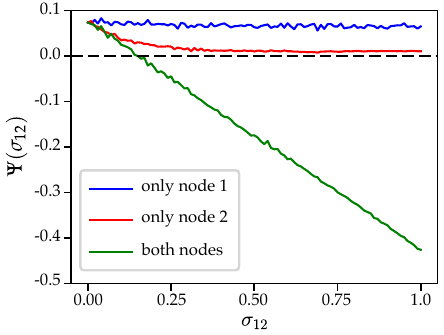}
    \caption{Master stability function $\psi(\sigma_{12})$ as a function of the coupling intensity $\sigma_{12}$ for connecting different mirrored oscillators with inter-layer links without switching 
    for the abovementioned parameters and $N=2$ and $\alpha=5$.  
    The blue curve represents the inter-layer link between mirrored oscillators 1, while the red curve represents the inter-layer link between mirrored oscillators 2. The green curve indicates the case where both inter-layer links are connected simultaneously. 
    }
    \label{fig:static}
\end{figure}

\begin{figure}[!htp]
    \centering
    \includegraphics[width=\columnwidth]{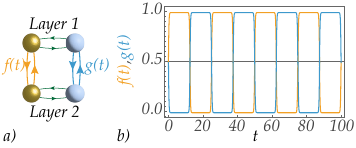}
    \caption{(a) Schematic of the two-layer system with two oscillators per layer, represented by blue and red circles. The intra-layer links are depicted in black. The inter-layer link connecting mirrored oscillators 1 is highlighted in red and mediated by the smooth square-wave function $f(t)$, while the inter-layer link connecting mirrored oscillators 2 is shown in blue and mediated by the smooth square-wave function $g(t)$. (b) The time evolution of the function $f(t)$ (red curve) and $g(t)$ (blue curve) is shown. The opposite phases between $f(t)$ and $g(t)$ illustrate the on/off dynamics of the inter-layer links. 
    The parameters used to illustrate smooth switching are $\alpha=5$ and $p=25$, i.e. $T_{swt}= 12.5$.
    }
    \label{fig:function}
\end{figure}

Now, to demonstrate that a single inter-layer link is insufficient to establish synchronization between the oscillators of the two-layer system, we calculate $\Psi(\sigma_{12}, T_{swt})$ as a function of $\sigma_{12}$ for the static case, i.e., without link switching. Hence, in Fig.~\ref{fig:static}, the blue curve illustrates the behavior of $\Psi(\sigma_{12})$ for the inter-layer link connecting oscillator number 1 from each layer (see schematic in Fig.~\ref{fig:function}(a) for the oscillators numbers). In this case, $\Psi(\sigma_{12})$ is positive within the considered interval of $\sigma_{12}$, indicating that the completely synchronized state is unstable for this inter-layer topology. Similarly, the red curve on the positive side of $\Psi(\sigma_{12})$ shows that the synchronized state is also unstable for the inter-layer link connecting oscillator number 2 from each layer. Finally, the green curve in Fig.~\ref{fig:static} represents the case in which both oscillators 1 and 2 from each layer receive an inter-layer link. In this case, $\Psi(\sigma_{12})$ becomes negative around $\sigma_{12} = 0.15$, indicating the onset of stable synchronized states. Therefore, stable complete synchronization among the oscillators from different layers is possible only when the number of inter-layer links equals the number of oscillators per layer.

Thus, we confirm that a two-layer system with two oscillators per layer cannot achieve complete synchronization when connected by only a single inter-layer link. Next, we would like to consider dynamically switching this link between the two oscillators in each layer. However, it is important to note that such link switching may eventually introduce  discontinuities in the system,  raising questions about the well-definedness and convergence of $\Psi(\sigma_{12}, T_{swt})$. 
To address this difficulty, we couple the oscillator via a smooth function that approximates a piecewise constant one, periodically alternating between $1$ and $0$ within prescribed time intervals. This function $g\left(t, \alpha, p \right)$ is defined as:
\begin{equation}
\begin{aligned}
\nonumber
g(t, \alpha, p) &= \tanh(\alpha(t - np)) \\
&- \tanh(\alpha(t - (n+0.5)p)) - 1,
\label{Eq:piecewise}
\end{aligned}
\end{equation}
where the parameter $\alpha$ controls the sharpness of the transition between $0$ and $1$, and $p$ determines the periodicity of $g(t, \alpha, p)$. The analysis of the synchronization errors shows that the results are coinciding for $\alpha\geq 5$. Therefore, $\alpha=5$ is used for all simulations. In this context, the period $p$ is twice the switching time, $T_{swt}$. The variable $n = \left\lfloor \frac{t}{p} \right\rfloor$ denotes the number of complete periods of length $p$ that fit within a given time interval. Consequently, $g\left(t, \alpha, p \right)$ generates a periodic pattern with intervals of length $p$, creating a smooth, square-wave-like shape. Therefore, this function offers a continuous method for switching an inter-layer link on and off, connecting a pair of mirror oscillators in a two-layer system containing two oscillators per layer. To simulate the switching of the inter-layer link between pairs of mirror oscillators, we connect one pair of mirror oscillators using a function $f\left(t, \alpha, p \right)= g(t+p,\alpha,p)$, which is in the opposite phase of $g\left(t, \alpha, p \right)$ connecting the other pair. The schematic shown in Fig.~\ref{fig:function}(a) illustrates the connection between mirror oscillators number 2 by the function $g\left(t, \alpha, p \right)$ in blue, while the connection between mirror oscillator number 1 is mediated by the function $f\left(t, \alpha, p \right)$ in red. In addition, in Fig.~\ref{fig:function}(b), we illustrate the behavior of $g\left(t, \alpha, p \right)$ and $f\left(t, \alpha, p \right)$, simulating the switching of the inter-layer link between the two mirror oscillators.

\begin{figure}[!htp]
    \centering
    \includegraphics[width=\columnwidth]{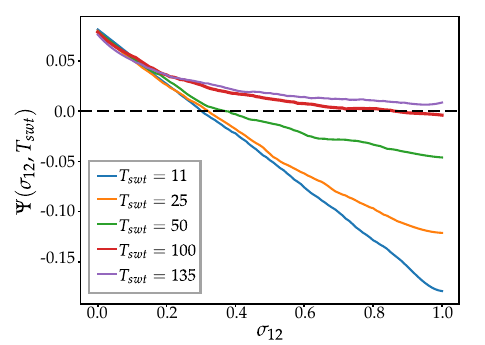}
    \caption{Master stability function $\Psi(\sigma_{12}, T_{swt})$ as function of the coupling intensity $\sigma_{12}$ and the switching time $T_{swt}$ 
    for the abovementioned parameters and $N=2$ and $\alpha=5$.
    The colored curves indicate different values of $T_{swt}$.     }
    \label{fig:switching}
\end{figure}

Next, we perform a transverse stability analysis using the MSF for the system with such continuous switching of the inter-layer link. Hence, in Fig.~\ref{fig:switching}, we present $\Psi(\sigma_{12})$ as a function of coupling intensity $\sigma_{12}$ for various switching times $T_{swt}$ (indicated by different colored curves). For $T_{swt}=135$ (purple curve in Fig.~\ref{fig:switching}), we observe that $\Psi(\sigma_{12})$ remains positive across the entire range of $\sigma_{12}$, indicating that stable, fully synchronized states do not occur, even at higher values of $\sigma_{12}$. However, at a shorter switching time, $T_{swt}=100$, we find that $\Psi(\sigma_{12})$ crosses the zero axis at approximately $\sigma_{12}=0.9$ (red curve in Fig.~\ref{fig:switching}), signaling the onset of stable, fully synchronized states. As we further decrease the values of $T_{swt}$, the onset of fully synchronized states occurs at progressively smaller values of $\sigma_{12}$. This is illustrated in Fig.~\ref{fig:switching}, where the intercepts of the green and yellow curves correspond to $T_{swt}=50$ and $T_{swt}=25$, respectively. Finally, for $T_{swt}=11$, the corresponding curve for $\Psi(\sigma_{12})$ crosses the zero axis at the smallest observed value of $\sigma_{12}$. Thus, decreasing $T_{swt}$ further does not facilitate the onset of fully synchronized states at even lower values of $\sigma_{12}$. In addition, by comparing the stability results in Fig.~\ref{fig:switching} with those in Fig.~\ref{fig:static} for the static inter-layer topology, we point out that fast switching in Fig.~\ref{fig:switching} effectively mimics the stability case in which both inter-layer links are connected in the static topology (green curve in Fig.~\ref{fig:static}), whereas slow switching tends toward the unstable case with only a single static inter-layer link (red and blue curves in Fig.~\ref{fig:static}).

\begin{figure}[!htp]
    \centering
    \includegraphics[width=\columnwidth]
    {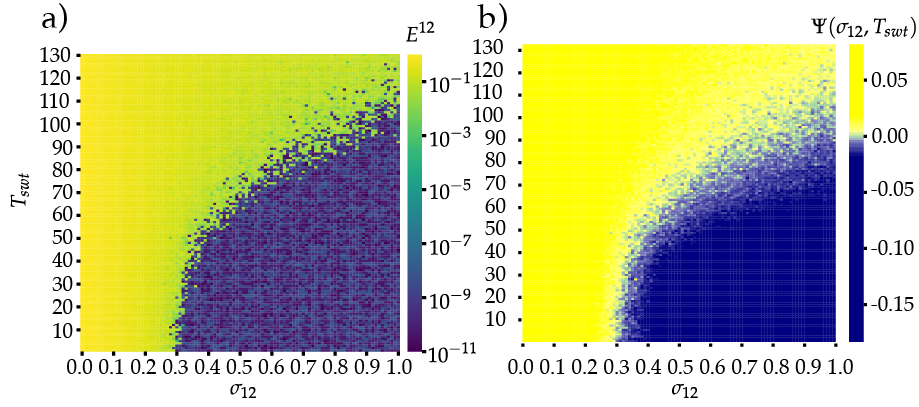}
    \caption{a) {The color code indicates the inter-layer synchronization error $E^{12}$, defined in Eq.~\eqref{eq:e12av}, as a function of the coupling intensity $\sigma_{12}$ and switching time $T_{swt}$. b) The colors indicates the master stability function $\Psi(\sigma_{12}, T_{swt})$ computed for $\sigma_{12}$ and $T_{swt}$ in the same intervals as in a). 
    The parameters are the above-mentioned and $N=2$ and $\alpha=5$.} 
 }
    \label{fig:error_switching}
\end{figure}

To clarify the interplay between the switching time $T_{swt}$ and the coupling intensity $\sigma_{12}$ in promoting stable synchronization within a sparsely coupled two-layer network, we calculate the inter-layer synchronization error $E^{12}$ and the master stability function $\Psi$ as functions of $T_{swt}$ and $\sigma_{12}$. The inter-layer synchronization error can be derived from its time-dependent version defined in Eq.~(\ref{eqn:td_error}) by computing the time average as 
\begin{equation}
E^{12} = \frac{1}{T} \int_{t}^{t+T} E(t') dt', \label{eq:e12av}
\end{equation} where $t = 1000$ and $T=10000$. Thus, in Fig.~\ref{fig:error_switching}(a), we show $E^{12}(T_{swt}, \sigma_{12})$ for 100 realizations of the two-layer system, with each layer containing two FHN oscillators, as described in Fig.~\ref{fig:function}(a). In this figure, the yellow regions mark well-defined parameter pairs ($T_{swt}$, $\sigma_{12}$) associated with high values of $E^{12}(T_{swt}, \sigma_{12})$, indicating desynchronization, while the dark-blue regions indicate ($T_{swt}$, $\sigma_{12}$) pairs that induce the system to complete synchronization. The dark-blue region, indicating induced synchronization, appears below a certain threshold for decreasing values of $T_{swt}$. Thus, the faster the links are switched across the network, the more easily it synchronizes -- requiring a lower coupling intensity for synchronization. Therefore, in the networks considered, reducing the time that oscillators remain connected, while increasing the number of connected oscillators, facilitates synchronization more effectively than extending the connection duration. In Fig.~\ref{fig:error_switching}(b), we present the master stability function $\Psi(\sigma_{12}, T_{swt})$ as a function of $T_{swt}$ and $\sigma_{12}$, within the same parameter range as in Fig.~\ref{fig:error_switching}(a). This stability analysis shows a strong correspondence with the synchronization error, confirming that the synchronous states emerging from network link switching are asymptotically stable.
In particular, we note that the stable region indicated by the negative values of the MSF in Fig.~\ref{fig:error_switching}(b) aligns well with the dark-blue low-error region in Fig.~\ref{fig:error_switching}(a). However, minor differences can appear due to finite-size effects and numerical integration errors.

Furthermore, the results in Figs.~\ref{fig:error_switching}(a) and \ref{fig:error_switching}(b) effectively illustrate the relationship between $T_{swt}$ and $\sigma_{12}$. Specifically, they clarify that below the switching time threshold of approximately $T_{swt} = 40$, reducing $T_{swt}$ does not enable synchronization at a coupling intensity lower than about $\sigma_{12} \approx 0.33$.

\section{Conclusions}

In summary, in a double-layer network of FitzHugh–Nagumo (FHN) oscillators, where each layer comprises two species of FHN oscillators exhibiting chaotic dynamics, we have numerically demonstrated that the switching of sparse inter-layer links can induce inter-layer synchronized states. More specifically, for a fixed inter-layer coupling intensity, we observe that when the interval between link switches is large, maintaining connections between mirror oscillators in each layer for extended durations is not effective in achieving complete inter-layer synchronization. Conversely, shorter intervals between link switches, in which each pair of mirrored oscillators remains connected for briefer periods, promote inter-layer synchronization more effectively. Subsequently, the effectiveness of shorter switching times is demonstrated by estimating the time required to achieve complete inter-layer synchronization for different layer sizes.

To gain insight into the transverse stability of the observed states of complete inter-layer synchronization induced by intermittently switching links in our double-layers networks, we apply the master stability function (MSF) formalism to a minimal double-layer system containing two oscillators per layer. We first note that the MSF analysis for networks with link switching presents inherent challenges due to the discontinuities that such switching may introduce into the system dynamics. To address this issue, we introduce smooth square-wave functions that emulate the link switching, thereby enabling a well-defined stability analysis. With this approach, we demonstrate that intermittently switching the links can induce stable synchronized states in the minimal system. This strategy captures the fundamental transverse stabilization mechanism induced by link switching.

We point out that, in our large double-layer system with sparse inter-layer topologies, synchronization is governed by local transverse modes associated with individual inter-layer connections, rather than by global collective coupling. Fast switching effectively generates a time-averaged stabilizing interaction, and each active inter-layer link locally enforces transverse contraction in the same manner as in the minimal system. Consequently, the stabilization mechanism identified for the minimal system constitutes a fundamental building block of the synchronization process in large sparse double-layer networks, thereby providing a sound justification for extending the minimal-system results on transversal stability to the large-scale regime considered here.

We emphasize that spontaneous intra-layer synchronization would reduce the effective layer dimensionality, thereby facilitating the onset of inter-layer synchronization through the switching strategy. Accordingly, the use of two oscillator populations within each layer represents a first step toward introducing heterogeneities that suppress spontaneous intra-layer synchronization. We further note that stronger dynamical heterogeneities at the oscillator level, or the introduction of topological heterogeneity such as random intra-layer connections, would pose additional difficulties for the establishment of inter-layer synchronization via link switching. Therefore, the investigation of different types of intra-layer heterogeneities is left for future work. As additional open questions for future exploration, we highlight the potential of applying switching connections to control the onset of stable collective states beyond synchronization, such as coherent motion in active systems \cite{Jadbabaie2003,Olfati2006,Su2009,Medeiros2024}, polarized opinion patterns in social groups \cite{Baumann2021,Perez2025}, and spreading patterns of epidemics \cite{Pastor2015,Guo2024}.

\begin{acknowledgments}
We want to thank Anna Zakharova for introducing us to this topic and her valuable discussions, which significantly enhanced this research. E.S.M acknowledges support from The S\~ao Paulo Research Foundation (FAPESP), project number 2023/15040-0.
M.E.~has been supported by Deutsche Forschungsgemeinschaft (DFG) through Germany's Excellence Strategy -- The Berlin Mathematics Research Center MATH+ EXC-2046/1, project 390685689, in particular subprojects AA1-8 and AA1-18. M.E. additionally thanks the Einstein Foundation 
and the NWO (VI.Vidi.233.133) 
for supporting his research.
\end{acknowledgments}

\bibliography{bib_adapt}

\end{document}